\documentclass[11pt]{article}

\usepackage{graphicx}                  
\usepackage{dcolumn}                   
\usepackage{bm}

\usepackage{jheppub}                   

\makeatletter
\gdef\@fpheader{\ }                    
\makeatother

\usepackage{graphicx}
\usepackage{dcolumn}
\usepackage{bm}
\usepackage{float}

\usepackage{amsmath,amssymb}           
\usepackage{amscd}                     
\usepackage{epsfig}                    
\usepackage[matrix,arrow]{xy}          
\usepackage{xspace}                    
\usepackage{stmaryrd}                  
\usepackage{slashed}
\usepackage{tikz}
\usepackage{hyperref}

\DeclareSymbolFont{bbold}{U}{bbold}{m}{n}
\DeclareSymbolFontAlphabet{\mathbbold}{bbold}

\newcommand{\bq}{\begin{equation}}
\newcommand{\eq}{\end{equation}}
\newcommand{\bea}{\begin{eqnarray}}
\newcommand{\eea}{\end{eqnarray}}

\newcommand{\dd}{\mathrm{d}}
\newcommand{\ee}{\mathrm{e}}

\newcommand{\der}{\partial}

\newcommand{\bbR}{\mathbb{R}}

\DeclareMathOperator{\SO}{\mathit{SO}}
\DeclareMathOperator{\SL}{\mathit{SL}}
\DeclareMathOperator{\GL}{\mathit{GL}}

\newcommand{\rep}[1]{\mathbf{#1}}

\newcommand{\id}{\mathbbold{1}}

\DeclareMathOperator{\diag}{diag}

\newcommand{\Dgen}{{D}}

\newcommand{\LC}{\nabla}

\newcommand{\proj}[1]{\times_{#1}}

\newcommand{\inn}{\mathbin{\lrcorner}}

\newcommand{\ba}{\bar{a}}
\newcommand{\bb}{\bar{b}}
\newcommand{\bc}{\bar{c}}

\newcommand{\tv}{\tilde{v}}

\newcommand{\hE}{\hat{E}}
\newcommand{\he}{\hat{e}}

\newcommand{\hs}[1]{\hspace{#1}}

\newcommand{\ra}{\rightarrow}

\newcommand{\Ra}{\Rightarrow}

\newcommand{\cH}{\mathcal{H}}
\newcommand{\tlambda}{\tilde{\lambda}}

\title{Classical worldvolumes as generalised geodesics}

\author{Charles Strickland-Constable}
\emailAdd{c.strickland-constable@herts.ac.uk}

\affiliation{Department of Physics, Astronomy and Mathematics, University of Hertfordshire, \\College Lane, Hatfield, AL10 9AB, UK}

\abstract{ 
It is a standard result that the integral curves of an auto-parallel vector field are geodesics which, for null and timelike vectors, are the paths of freely-falling particles in general relativity. We introduce a definition of an ``auto-parallel" generalised vector field and show that it gives the analogous statements for the classical worldvolumes of strings and branes in arbitrary background field configurations. This appears to give a unified description of the worldvolume equations of strings and branes, similar to the way that generalised geometry provides a unified description of maximal supergravity theories. We present details of the cases of string worldsheets in $O(10,10)$ generalised geometry and M2 branes restricted to the four dimensions of $\SL(5,\bbR)\times\bbR^+$ generalised geometry. A key quantity is the infinitesimal flow of the conjugate momentum along the generalised tangent vector, which is equated to the gradient of the Hamiltonian, viewed as a function on spacetime. 
\vfill}

\begin{document}  
 
\maketitle



\section{Introduction}

\label{sec:intro}

In ordinary differential geometry, a vector field $X$ defines a congruence of (its integral) curves. 
In standard presentations, it is stated that if the vector field is auto-parallel in the Levi-Civita connection (i.e. it is parallel-transported along its integral curves)
\begin{equation}
\label{eq:ordinary-geo}
	\LC_X X = 0 \;,
\end{equation}
then these curves are (affinely-parameterised) geodesics for the corresponding metric. In general relativity (GR), the metric has Lorentzian signature and if these curves are timelike or null, then they describe the trajectories of massive or massless particles freely-falling in the gravitational field encoded by the metric (see e.g.~\cite{Wald}). In the timelike case, these geodesic paths minimise the action functional given by the invariant relativistic length of the path, given the boundary conditions. An action which describes also null geodesics can be found by introducing a worldline metric. Using worldline reparameterisation invariance, this can be set equal to one, leaving the Lagrangian as the square of the invariant interval. We will consider this gauge-fixed action and its generalisations in this work. 

In string theory, the analogue of classical freely-falling particles in GR are classical strings (see e.g.~\cite{GSWvol1,Polchinski1,BBS}). Their dynamics are described by a mathematically similar action to the point particle, but generalised to the case where the worldline has become a two-dimensional worldsheet. In addition to worldsheet reparameterisation invariance, this is invariant under local rescalings of the worldsheet metric, which enables us to write it in a gauge-fixed form very similar to that of the particle. There is also a coupling to a two-form potential $B$ in the target space, given simply as the integral of the pull-back of $B$ to the worldsheet. Similar statements can be made about other branes arising in string theories, albeit with additional complications due to non-linearities and the absence of conformal symmetry. 

In classical GR, the study of geodesics is fundamental to understanding the structure of the theory and forms a key component of all aspects from orbits of planets to singularity theorems. In string theory, historically the focus has largely been centred on questions of quantisation, but understanding classical string (and brane) solutions has been of significant interest, for example, in semi-classical analysis of the AdS/CFT correspondence~\cite{Berenstein:2002jq,Gubser:2002tv,Russo:2002sr,Sezgin:2002rt,Mandal:2002fs,Alishahiha:2002sy,Minahan:2002rc,Alishahiha:2002fi,Frolov:2003qc,Frolov:2003xy,Tseytlin:2003ii,Bobev:2005cz,Hartnoll:2002th}. 

Given the similarities (at least in spirit) between the mathematical descriptions of all of these objects, it is natural to wonder if there is an analogue of the tangent vector $X$ for, say, strings. Further, one can also consider what might be the analogue of equation~\eqref{eq:ordinary-geo} in the case that one extends the tangent vector to be a field over spacetime. Ideally, we would like an equation such that the tangent vector object defines a foliation of spacetime by string worldsheets if it satisfies the differential equation. 

At first glance, one would say that a two-dimensional surface will have two ordinary tangent vectors, and the equations of motion become some differential equations relating them to the background metric and field strength $H =\dd B$. One could leave it at that. However, it has been noted long ago~\cite{Gates:1984nk} that these equations can be viewed as the preservation of the tangent vectors in the worldsheet-null directions by the connections with torsion $\LC^{(\pm)} \sim \LC \pm \tfrac12 H$, suggesting that they have more geometric structure. In this work we will show that these equations have considerably more structure still, which generalises also to other branes. 

As the target space ingredients of the string equations are the metric and $B$ field, one immediately suspects that generalised geometry~\cite{Hitchin,Gualtieri}, which combines these objects into a generalised metric, will be a suitable framework in which to look for such additional structure. While this is indeed the framework used in this paper, we should note that there is already a wide literature studying how generalised geometry, and similar constructions using a doubled spacetime, describe supergravity, string worldsheets and non-geometric backgrounds~\cite{
Duff:1989tf,Tseytlin:1990nb,Tseytlin:1990va,Hull:2004in,Hull:2006va,Berman:2007xn,Hohm:2013jaa,Blair:2013noa,Arvanitakis:2018hfn,Bonezzi:2020ryb,Siegel:1993th,Hull:2009mi,Hohm:2010pp,Lindstrom:2004eh,Lindstrom:2004iw,Lindstrom:2005zr,CSW1}. As far as worldsheet statements are concerned, many of these works have focused on either the geometry of the target space of sigma-models with supersymmetry or the construction of actions for strings and branes, looking to make the duality symmetries manifest in the formulation and to quantise the systems in those terms.

In this work, we will encounter several ideas which have inevitably appeared before in this literature. However, we will focus only on the classical equations of motion of the objects, rather than action principles. We see how the two ordinary tangent vectors $v = \tfrac{\der}{\der \tau}$ and $\tv = \tfrac{\der}{\der \sigma}$ of a string worldsheet are combined into a generalised vector $V = v + g \cdot \tv$. 
The Virasoro constraints are then the vanishing bilinears of this vector in the (Lorentzian signature) generalised metric $G$ and the $O(10,10)$ metric on the generalised tangent space. 
We then imagine that this is extended off the worldsheet of a single string to a vector field on spacetime, in at least some open set containing the worldsheet, similarly to how one can move from the tangent vector of a curve to a local vector field $X$. This is so that we can define derivatives of the vector field in all directions in spacetime, rather than purely along the worldsheet, even though in the end these must cancel out from the equations. In this way, we will formulate our discussion in terms of a generalised vector field on the target space. 

Our key results will concern the formulation of the equations of motion in terms of this generalised vector $V$. These equations, together with the Virasoro conditions, then encapsulate the full system in our language. If they are satisfied on a patch of the target spacetime, then we have a foliation of spacetime by classical string worldsheet solutions. (Other results on the existence of such foliations have appeared in~\cite{Severa,Chatzistavrakidis:2016jfz}). However, one could also simply require them to be solved on a single two-dimensional worldsheet (such that $V$ is the generalised tangent vector to it) and this would then give an isolated string worldsheet solution. As the derivatives which are not along the worldsheet cancel from the equations, the manner in which the generalised tangent vector is extended to a local field does not affect them. 

As a first pass, we write the equation of motion as two equations, each resembling~\eqref{eq:ordinary-geo}, but using generalised Levi-Civita connections. These are simply the equations in terms of $\LC^{(\pm)}$ from~\cite{Gates:1984nk} re-branded as generalised geometry objects. 

However, we then reformulate the ordinary geodesic equation~\eqref{eq:ordinary-geo} in a form utilising the Lie derivative, such that no connection appears explicitly. We later interpret this in terms of the conjugate momentum and Hamiltonian viewed as a function on spacetime. Remarkably, writing the exact same equation using the Dorfman derivative (or generalised Lie derivative) and generalised metric we recover the equations of motion for the string. We thus propose this as a plausible definition of the analogue of the auto-parallel condition for a generalised vector, even though no notion of generalised parallel transport itself is developed here. The objects in this equation have the same interpretation as for the ordinary geodesic equation, but for a generalised covector conjugate momentum, similar to those which have been discussed in previous studies (see e.g.~\cite{Blair:2013noa,Bonezzi:2020ryb}). It is also noteworthy that our condition reproduces the equation of motion without using the quadratic constraints on the generalised tangent vector, suggesting that it could be a meaningful definition also in the absence of these constraints (as~\eqref{eq:ordinary-geo} is also for non-null vector fields). 

Further to the above, we show that the exactly analogous equations written in a different form of generalised geometry, with generalised structure group $\SL(5,\bbR)\times\bbR^+$, describe the equations for the M2 brane restricted to the four dimensions (out of eleven) which are included in this generalised geometry. This is the four-dimensional case of exceptional generalised geometry~\cite{chris,PW} which describes the dimensional restrictions of eleven-dimensional supergravity (and type II supergravity via different decompositions)~\cite{CSW2,CSW3}, and forms the basis for the internal part of exceptional field theory~\cite{HS1,HS2,Godazgar:2014nqa}. As for the string, we use a gauge-fixed form of the M2 brane theory, with the gauge fixing constraints now corresponding to the quadratic constraints on our generalised tangent vector. 
In the process we encounter objects reminiscent of those in previous studies of membrane worldvolume theories in the context of extended geometry~\cite{Duff:1990hn,BermanPerry,Duff:2015jka,Hatsuda:2012vm,Sakatani:2016sko,Sakatani:2017vbd}. 
Our construction then proceeds in exactly the same way as in our discussion of the string. This suggests that our definition of the auto-parallel condition and formulation of the equations of motion will be universal across generalised geometries. We also stress that no assumptions are made about the nature of the background fields, which need not solve the supergravity equations of motion. 

The structure of the paper is as follows. In section~\ref{sec:strings} we review the action and equations of motion for the string in a background metric and $B$-field, introduce the necessary elements of $O(10,10)$ generalised geometry and show how to formulate the equations of motion using a generalised Levi-Civita connection. Next, in section~\ref{sec:auto-parallel} we reformulate the ordinary auto-parallel condition without explicit use of a connection and give our definition of an auto-parallel generalised vector field. This is then shown to reproduce the equations of section~\ref{sec:strings} for the string in $O(10,10)$ generalised geometry. In section~\ref{sec:M2} we provide the corresponding statements for the M2 brane in $\SL(5,\bbR)\times\bbR^+$ generalised geometry. We end with some discussion of our results in section~\ref{sec:discussion}.


\section{Classical strings}
\label{sec:strings}


\subsection{Action and equations of motion}

The action of the classical string in a background field configuration $(g,B)$, gauge fixed to conformal gauge on the worldsheet, is
\begin{equation}
\label{eq:string-action}
	S = -\frac12 \int \dd^2 s \Big(\eta^{\alpha\beta} g_{mn} + \epsilon^{\alpha\beta} B_{mn} \Big) \frac{\der x^m}{\der s^\alpha} \frac{\der x^n}{\der s^\beta}
\end{equation}
The resulting classical equations of motion can be written as
\begin{equation}
	\eta^{\alpha\beta} \Big[ \frac{\der^2 x^m}{\der s^\alpha \der s^\beta} 
		+ \Gamma_p{}^m{}_q \frac{\der x^p}{\der s^\alpha} \frac{\der x^q}{\der s^\beta} \Big]
		- \frac12 \epsilon^{\alpha\beta} H^m{}_{pq} 
			\frac{\der x^p}{\der s^\alpha} \frac{\der x^q}{\der s^\beta} = 0
\end{equation}
Writing this out explicitly using $\eta = \diag(-1,+1)$ and $(s^\alpha) = (\tau,\sigma)$ and denoting 
\begin{equation}
	v^m = \frac{\der x^m}{\der \tau}
	\qquad 
	\tv^m = \frac{\der x^m}{\der \sigma}
\end{equation}
we have\footnote{Here we take $\epsilon^{01} = +1$. This matches the sign conventions of the $B$-field in the generalised geometry construction.}
\begin{equation}
	- \frac{D v^m}{D \tau} + \frac{D \tv^m}{D \sigma} - H^m{}_{pq} v^p \tv^q = 0
\end{equation}
where for any vector $w \in \Gamma(TM|_{\text{worldsheet}})$
\begin{equation}
	\frac{D w^m}{D \tau} = \frac{\der w^m}{\der \tau} 
		+ v^p \Gamma_p{}^m{}_n w^n
	\qquad 
	\frac{D w^m}{D \sigma} = \frac{\der w^m}{\der \sigma} 
		+ \tv^p \Gamma_p{}^m{}_n w^n
\end{equation}
are the target space covariant derivatives along the $\tau$ and $\sigma$ directions. 

In addition to these equations of motion, one must also impose the Virasoro constraints, which are the equations of motion of the worldsheet metric written in the conformal gauge in which we wrote our action~\eqref{eq:string-action}. These are the vanishing of the energy momentum tensor
\begin{equation}
	T_{\alpha \beta} = g_{mn} \der_\alpha x^m \der_\beta x^n 
		- \tfrac12 \eta_{\alpha \beta} \eta^{\gamma \delta} g_{mn} \der_\gamma x^m \der_\delta x^n
\end{equation}

Let us now extend to the analogue of a congruence of curves, and consider a coordinate system on spacetime for which $\tau$ and $\sigma$ are the first two coordinates (often called ``static gauge" in the literature). 
We then promote $v^m$ and $\tv^m$ to vector fields on a patch of spacetime, rather than just on a two-dimensional embedded worldsheet. If these satisfy the relevant equation of motion they will then define a foliation of the patch of spacetime by string worldsheets. 

In that setup, the above equation of motion becomes
\begin{equation}
\label{eq:string-eom-vtv}
	- v^p \LC_p v^m + \tv^p \LC_p \tv^m - H^m{}_{pq} v^p \tv^q = 0
\end{equation}
while the Virasoro constraints are
\begin{equation}
\label{eq:vector-Virasoro}
	g(v,v) + g(\tv, \tv) = 0
	\qquad
	g(v,\tv) = 0
\end{equation}
Clearly, it is possible to satisfy the first of~\eqref{eq:vector-Virasoro} with both $v$ and $\tv$ non-zero only if the metric $g$ has indefinite signature.

In section~\ref{sec:gen-geo-1} we will recover this system from generalised geometry via generalised connections. In section~\ref{sec:gen-string-2} we will see a more universal generalised geometry formulation.


\subsection{Elements of $O(10,10)$ generalised geometry}
\label{sec:gen-geom-Odd}

In this section, we briefly recall some of the features of generalised geometry needed to describe the string. 
We mostly follow the presentation of~\cite{CSW1}, to which the reader can refer for full details of these constructions. 

Firstly, our spacetime is equipped with a $B$-field, which may not be globally defined, so we introduce patches of the spacetime such that between the patches the $B$-field transforms as
\begin{equation}
	B' = B - \dd \Lambda
\end{equation}
A generalised vector field $V = v+\lambda$ is a vector field $v$ together with a one-form $\lambda$ defined on each of the patches as above. Between the patches, the one-form part transforms as
\begin{equation}
	\lambda' = \lambda - i_v \dd \Lambda
\end{equation}
As these one-forms are explicitly twisted by the gauge transformations in this way, we will refer to this as the ``twisted picture" representation of a generalised vector. The action of these gauge transformations preserves the $O(10,10)$ inner product $\eta$
\begin{equation}
	\langle V, V \rangle = \eta(V,V) = i_v \lambda
\end{equation}
and so, including also the $\GL(10,\bbR)$ action of diffeomorphisms, we can think of the structure group of the generalised tangent space to be $O(10,10)$ (even though in fact it lies only in a parabolic subgroup).

One can also discuss the ``untwisted picture" representation of generalised vectors. The generalised tangent bundle $E$ is isomorphic to the direct sum $T \oplus T^*$, and the isomorphism can be made explicit using the $B$-field. The one-form
\begin{equation}
	\tlambda = \lambda - i_v B
\end{equation}
is invariant under the gauge transformations between patches, and so $v + \tlambda$ is a well-defined section of $T \oplus T^*$ over our spacetime.

In~\cite{CSW1}, this isomorphism is presented in terms of the components of generalised vectors with respect to certain split frames for the generalised tangent space. Further discussion of the twisted vs untwisted pictures can be found in~\cite{CdFPSW}. In this paper, we will do most of our calculations working with the untwisted picture representation, so that generalised vectors will simply be the sum of a vector field and a one-form field (and we will drop the tilde on the one-form from the notation). 

One of the key structures in generalised geometry is the Dorfman derivative (or generalised Lie derivative) which generates the action of infinitesimal generalised diffeomorphisms (i.e. combined diffeomorphisms and $B$-field gauge transformations). With respect to the twisted picture components, we have 
\begin{equation}
	L_V V' = [v,v'] + \mathcal{L}_v \lambda' - i_v \dd \lambda
\end{equation}
Introducing $O(10,10)$ indices $M, N$ via
\begin{equation}
\label{eq:Odd-indices}
	\big(V^M\big) = \begin{pmatrix} v^m \\ \lambda_m \end{pmatrix}
	\hs{30pt}
	\big(\der_M\big) = \begin{pmatrix} \der_m \\ 0 \end{pmatrix}
	\hs{30pt}
	\big(\eta_{MN} \big) = \frac12 \begin{pmatrix} 0 & \id \\ \id & 0 \end{pmatrix}
\end{equation}
and raising and lowering these indices with $\eta$ and its inverse, we can write the Dorfman derivative as
\begin{equation}
	(L_V V')^M = V^N \der_N V'^M + (\der^M V_N - \der_N V^M) V'^N
\end{equation}
Its action on an section $W$ of $E^*$, written with a lower index, is then
\begin{equation}
\label{eq:Dorfman-Estar}
	(L_V W)_M = V^N \der_N W_M + (\der_M V^N - \der^N V_M) W_N
\end{equation}

In the untwisted picture, this takes the form
\begin{equation}
	L_V V' = [v,v'] + \mathcal{L}_v \lambda' - i_v \dd \lambda - i_v i_{v'} H
\end{equation}
where $\mathcal{L}_v$ denotes the ordinary Lie derivative and $H =\dd B$ is the field strength of $B$. This satisfies the Leibniz identity, giving $E$ the structure of a Leibniz algebroid~\cite{Baraglia}, though it is usually referred to as a Courant algebroid~\cite{roytenberg}, as it has more structure still. It is not a Lie algebroid as the Dorfman derivative is not anti-symmetric, but the symmetric part is exact
\begin{equation}
\label{eq:Dorfman-sym}
	L_V V' + L_{V'} V = 2 \dd \langle V, V' \rangle
\end{equation}

Another important object for us will be the generalised metric $G$, which gives another inner product on $E$.\footnote{Note again that when writing $O(10,10)$ indices, we will use the $O(10,10)$ inner product $\eta$ to raise and lower them.} In the untwisted picture, the generalised metric can be written simply in terms of a metric $g$ on the spacetime via
\begin{equation}
	G(V,V) = \tfrac12 \Big[ g(v,v) + g^{-1}(\lambda, \lambda) \Big]
\end{equation}
In this paper, we will take the spacetime metric $g$ to have signature $(9,1)$ so that $G$ is stabilised by $O(9,1)\times O(9,1) \subset O(10,10)$. 

One can also introduce frames for the generalised tangent space which diagonalise both $\eta$ and $G$. Given two orthonormal frames $\he^+_a$ and $\he^-_{\ba}$ for the tangent bundle (with duals $e^{+a}$ and $e^{-\ba}$) these can be defined (in the untwisted picture) by
\begin{equation}
\label{eq:gen-frames}
\begin{aligned}
	\hE^+_a &= \he^+_a + e^+_a
	\\
	\hE^-_{\ba} &= \he^-_{\ba} - e^-_{\ba}
\end{aligned}
\end{equation}
Clearly, there is an $O(9,1)\times O(9,1)$ family of these frames rotating the $a$ and $\ba$ indices separately, reflecting the $O(9,1)\times O(9,1) \subset O(10,10)$ structure defined by the generalised metric. A generalised vector written with respect to these frames as $V = V^{+a} \hE^+_a + V^{-\ba}\hE^-_{\ba}$ has 
\begin{equation}
	G(V,V) = V^{+a} V^{+}_a + V^{-\ba} V^{-}_{\ba}
	\hs{30pt}
	\eta(V,V) = V^{+a} V^{+}_a - V^{-\ba} V^{-}_{\ba}
\end{equation}
Note that we raise and lower $a,b,c,\dots$ with $g_{ab}, g^{ab}$ and $\ba,\bb,\bc, \dots$ with $g_{\ba\bb}, g^{\ba\bb}$.\footnote{This can lead to clashes with signs when decomposing $O(10,10)$ indices. We must be careful to define that $V^{\ba} = V^{A=\ba}$ with an upper index and $D_{\ba} = D_{A=\ba}$ with a lower index.}

One can also define generalised connections acting on generalised tensor bundles $Q$ to be linear differential operators
\begin{equation}
	D : Q \longrightarrow E^* \otimes Q
\end{equation}
with a natural notion of generalised torsion
\begin{equation}
	L^{(D)}_V  - L_V  = T(V) \cdot 
\end{equation}
defined as the change in the Dorfman derivative when one inserts the generalised connection in place of the partial derivative. 

In~\cite{CSW1}, there is a lengthy discussion of how one can construct torsion-free generalised connections and in particular those which preserve the generalised metric. In fact, this involves introducing an auxiliary line bundle with structure group $\bbR^+$, so that the generalised structure group is enhanced to $O(10,10)\times\bbR^+$. The generalised Levi-Civita connections are then those which are compatible with an $\SO(9,1)\times\SO(9,1)$ subgroup of this, with the additional compatibility requirements introducing the dilaton field into the structure. There is no unique choice for any given generalised metric, but rather a family of such connections, though the undetermined parts drop out of all physical operators which are built from them. We will not recount the full construction here, but merely note the result that such generalised connections act on generalised vectors according to
\begin{equation}
\label{eq:Dgen-sol}
\begin{aligned} 
   \Dgen_a V_+^b 
       &= \nabla_a V_+^b - \tfrac{1}{6}H_a{}^b{}_cV_+^c
           - \tfrac{2}{9}\big( 
              \delta_a{}^b \der_c\phi-\eta_{ac}\der^b\phi \big)V_+^c 
           + A^+_a{}^b{}_c V_+^c , \\
   \Dgen_{\bar{a}} V_+^b 
       &= \nabla_{\bar{a}} V_+^b - \tfrac{1}{2}H_{\bar{a}}{}^b{}_cV_+^c , \\ 
   \Dgen_a V_-^{\bar{b}} 
       &= \nabla_a V_-^{\bar{b}} 
           + \tfrac{1}{2}H_a{}^{\bar{b}}{}_{\bar{c}}V_-^{\bar{c}} , \\
   \Dgen_{\bar{a}} V_-^{\bar{b}} 
       &= \nabla_{\bar{a}} V_-^{\bar{b}} 
           + \tfrac{1}{6}H_{\bar{a}}{}^{\bar{b}}{}_{\bar{c}}V_-^{\bar{c}}
           - \tfrac{2}{9}\big( 
              \delta_{\bar{a}}{}^{\bar{b}} \der_{\bar{c}}\phi
              - \eta_{\bar{a}\bar{c}}\der^{\bar{b}}\phi \big)V_-^{\bar{c}} 
           + A^-_{\bar{a}}{}^{\bar{b}}{}_{\bar{c}} V_-^{\bar{c}} , 
\end{aligned}
\end{equation}
where $A^\pm$ are undetermined tensors which have the symmetries
\begin{equation}
\label{eq:Adef}
\begin{aligned} 
   A^+_{abc}  &= -A^+_{acb} , & 
   A^+_{[abc]} &= 0 , &
   A^+_a{}^a{}_b &= 0 , \\
   A^-_{\bar{a}\bar{b}\bar{c}}  &= -A^-_{\bar{a}\bar{c}\bar{b}}, & 
   A^-_{[\bar{a}\bar{b}\bar{c}]} &= 0 , &
   A^-_{\bar{a}}{}^{\bar{a}}{}_{\bar{b}} &= 0 ,  
\end{aligned}
\end{equation}
%


\subsection{Classical strings as generalised geodesics 1: generalised connection approach}
\label{sec:gen-geo-1}

We form a generalised vector, using the $C_\pm$ bases $\hE^+_a$ and $\hE^-_{\ba}$ as 
\begin{equation}
\label{eq:Vpm}
	V^+{}^a  = \tfrac12 (v^a + \tv^a)
	\qquad 
	V^-{}^{\ba}  = \tfrac12 (v^{\ba} - \tv^{\ba})
\end{equation}
where for now, $v$ and $\tv$ are arbitrary vector fields. 
In what follows though, $v \pm \tv$ will be seen to be the null directions in the string worldsheet metric, making concrete the notion that the $V^\pm$ correspond to the left and right moving directions in the string, as remarked in previous works (e.g.~\cite{CSW1}).

We claim that the system of section~\ref{sec:strings} is encapsulated in the equations
\begin{equation}
\label{eq:gen-geo1}
	V^+{}^a D_a V^-{}^{\ba} =0
	\qquad 
	V^-{}^{\ba} D_{\ba} V^+{}^a = 0
\end{equation}
To see this, first we expand the equations using the explicit formulae for the generalised Levi-Civita connection~\eqref{eq:Dgen-sol}
\begin{equation}
\label{eq:gen-nabla-pm}
\begin{aligned}
	V^+{}^c D_c V^-{}^{\ba} &= V^+{}^c \Big( \LC_c V^-{}^{\ba} 
		+ \tfrac12 H_c{}^{\ba}{}_{\bb} V^-{} ^{\bb}\Big)
	\\
	V^-{}^{\bc} D_{\bc} V^+{}^a &= V^-{}^{\bc} \Big( \LC_{\bc} V^+{}^a  
		- \tfrac12 H_{\bc}{}^{a}{}_{b} V^+{}^b \Big)
\end{aligned}
\end{equation}
Now we use the components~\eqref{eq:Vpm} and align the frames $\he^+_a = \he^-_a = \he_a $ so as to effectively decompose under the Lorentz group of the tangent bundle which sits diagonally inside the generalised structure group $\SO(9,1)\times\SO(9,1)$. We find
\begin{equation}
\begin{aligned}
	4V^+{}^c \Big( \LC_c V^-{}^{a} 
		+ \tfrac12 H_c{}^{a}{}_{b} V^-{} ^{b}\Big)
	&= [\LC_v v - \LC_{\tv} \tv]^a + [\LC_{\tv} v - \LC_v \tv]^a \\
		& \qquad + \tfrac12 H_c{}^{a}{}_{b} (v^c v^b - \tv^c \tv^b + \tv^c v^b - v^c \tv^b) \\
	&= \Big( [\LC_v v - \LC_{\tv} \tv]^a +  H_c{}^{a}{}_{b} \tv^c v^b \Big) + [\LC_{\tv} v - \LC_v \tv]^a
		  \\
	4V^-{}^{\bc} \Big( \LC_{\bc} V^+{}^a  
		- \tfrac12 H_{\bc}{}^{a}{}_{b} V^+{}^b \Big) 
	&= [\LC_v v - \LC_{\tv} \tv]^a - [\LC_{\tv} v - \LC_v \tv]^a \\
		& \qquad - \tfrac12 H_c{}^{a}{}_{b} (v^c v^b - \tv^c \tv^b - \tv^c v^b + v^c \tv^b) \\
	&= \Big( [\LC_v v - \LC_{\tv} \tv]^a +  H_c{}^{a}{}_{b} \tv^c v^b \Big) - [\LC_{\tv} v - \LC_v \tv]^a
		  \\
\end{aligned}
\end{equation}
Thus we see that equations~\eqref{eq:gen-geo1} are equivalent to the pair of equations
\begin{equation}
	[\LC_v v - \LC_{\tv} \tv]^a +  H_c{}^{a}{}_{b} \tv^c v^b =0
	\qquad
	\LC_{\tv} v - \LC_v \tv = 0
\end{equation}
The first of these equations is the string equation of motion~\eqref{eq:string-eom-vtv}. The second equation looks like an unwanted additional condition. However, because the Levi-Civita connection is torsion-free this equation can be written as
\begin{equation}
	[v, \tv] = 0
\end{equation}
which is the condition that there are coordinates $\tau$ and $\sigma$ for which 
\begin{equation}
	v = \frac{\der}{\der \tau}
	\qquad
	\tv = \frac{\der}{\der \sigma}
\end{equation}
Thus equations~\eqref{eq:gen-geo1} on a generalised vector $V$ firstly impose that the constituent vectors $v$ and $\tv$ commute and thus define a foliation of spacetime by string worldsheets, and then also impose that those worldsheets are solutions of the classical equations of motion for the string in the background generalised metric $G$.

The Virasoro constraints~\eqref{eq:vector-Virasoro} are also neatly encapsulated in terms of the generalised metric and the $O(10,10)$ metric. In particular, 
\begin{equation}
\label{eq:gen-Virasoro}
\begin{aligned}
	G(V,V) &=  V^+{}^c V^+{}_c + V^-{}^{\bc} V^-{}_{\bc} 
		= \tfrac12 \big( g(v,v) + g(\tv,\tv) \big) \\
	\eta(V,V) &=   V^+{}^c V^+{}_c - V^-{}^{\bc} V^-{}_{\bc} =  g(v,\tv) \\
\end{aligned}
\end{equation}
are the two quantities which are set to zero by these. We thus conclude that a generalised vector field with non-vanishing $v$ and $\tv$, satisfying~\eqref{eq:gen-geo1}, with vanishing quantities~\eqref{eq:gen-Virasoro}, encodes a foliation of spacetime by classical string worldsheets. This is analogous to a null geodesic congruence in general relativity. We will come back to the issue of whether the worldsheet is non-singular, in the sense that both $v$ and $\tv$ are non-zero, in the discussion in the final section.

If one considers $V^{\pm a} \hat{e}_a$ as ordinary vector fields, the quantities appearing in equation~\eqref{eq:gen-nabla-pm} are of course the ordinary connections with torsion usually denoted by $\nabla^{\pm}$ in the literature. It has long been known~\cite{Gates:1984nk} that the equations of motion of the string could be written as the vanishing of the ordinary covariant derivatives in~\eqref{eq:gen-nabla-pm}, though the statement that given such vector fields $V^{\pm}$ one obtains a foliation of spacetime by string worldsheets (at least locally) has not been greatly emphasised. 

The conditions~\eqref{eq:gen-geo1} at first appear slightly ad hoc, and are not obviously recognisable as the analogue of the usual auto-parallel condition $\nabla_X X = 0$. In the next section, we will see that if one reformulates the ordinary auto-parallel condition suitably, equations~\eqref{eq:gen-geo1} are in fact precisely the analogous conditions in generalised geometry.


\section{The auto-parallel condition revisited}
\label{sec:auto-parallel}

We begin this section by writing a connection-free expression for the quantity $\LC_X X$, as this expression is the one which we will generalise in our key definition.  
By substituting in the Levi-Civita connection (using its torsion-free property so that $\mathcal{L} = \mathcal{L}^\nabla$ and $\dd = \dd_\nabla$), one can easily verify the identity
\begin{equation}
\label{eq:new-auto-parallel}
	\nabla_X X = g^{-1} \cdot \Big[ \mathcal{L}_X (g\cdot X) - \tfrac12 \dd [g(X,X)] \Big]
\end{equation}
for any vector field $X$, where we employ the notation that given a tangent vector $X$, $g \cdot X$ is the one-form obtained via $(g \cdot X)_m = g_{mn} X^n$.


\subsection{Classical strings as generalised geodesics 2: connection-free approach}
\label{sec:gen-string-2}

Correspondingly, we define that a generalised vector field $V$ is auto-parallel if
\begin{equation}
\label{eq:gen-auto-parallel}
	\mathcal{P}_V V := G^{-1} \cdot \Big[ L_V (G\cdot V) - \tfrac12 \dd [G(V,V)] \Big] = 0
\end{equation}
where $G$ is the generalised metric and $(G\cdot V)_M = G_{MN} V^N$ gives a section of $E^*$. 
We claim that the vanishing of this operator is precisely the conditions~\eqref{eq:gen-geo1}. We first demonstrate that this recovers equations~\eqref{eq:gen-geo1} by evaluating the expression in terms of generalised Levi-Civita connections. We then do an alternative calculation which does not involve generalised connections to recover equation~\eqref{eq:string-eom-vtv} directly. 
We note again that we have not defined a notion of parallel transport here, but choose the label auto-parallel as our expression precisely mirrors~\eqref{eq:new-auto-parallel}. 
This is similar to the philosophy by which integrable generalised $G$-structures were said to have generalised special holonomy in~\cite{CSW4}. 

Into the definition we can then insert any generalised Levi-Civita connection, since these are generalised torsion-free, and decompose into $O(9,1)\times O(9,1)$ objects. We have, for $W = G\cdot V$, using~\eqref{eq:Dorfman-Estar}
\begin{equation}
\begin{aligned}
	(L_V (G\cdot V))_a  
		& = V^B D_B W_a + (D_a V^B) W_B - \eta_{aa'} \eta^{BB'} (D_{B'} V^{a'}) W_B  \\
		&= V^b D_b W_a + V^{\bb} D_{\bb} W_a + (D_a V^b) W_b + (D_a V^{\bb}) W_{\bb}
			- W^b D_b V_a + W^{\bb} D_{\bb} V_a \\
		& = 2V^{\ba}  D_{\ba} V_a + V^c  D_a  V_c  + V^{\ba} D_a V_{\ba} \\
		& = 2V^{\ba}  D_{\ba} V_a + \tfrac12 D_a \big(G(V,V) \big)
\end{aligned}
\end{equation}
Next, viewing $\dd G(V,V)$ as a section of $E^*$\footnote{As the identification of $E$ with $E^*$ is via $\eta_{MN}$, and~\eqref{eq:Odd-indices} contains a factor $1/2$, there are many awkward factors of $2$ in this section. We could simply ignore the distinction between $E$ and $E^*$ here, but as we cannot do this in other generalised geometries, we maintain it.}, so that its inner product with a generalised vector $V' = v'+\lambda'$ is $v'^m \der_m G(V,V)$, we have 
\begin{equation}
\label{eq:dG-Estar}
\begin{aligned}[]
	[ \dd G(V,V)]_a  = \der_a (V^c  V_c  + V^{\bc}V_{\bc} ) = \der_a \big(G(V,V) \big)
\end{aligned}
\end{equation}
so that 
\begin{equation}
	(\mathcal{P}_V V)^a = \Big[ G^{-1} \cdot \big[ L_V (G\cdot V) - \tfrac12 \dd [G(V,V)] \big] \Big]^a 
	= 2 V^{\ba}  D_{\ba} V^a
\end{equation}
is (twice) the operator in the first of equations~\eqref{eq:gen-geo1}. 
Similarly, keeping careful track of conventional factors, we find
\begin{equation}
	(\mathcal{P}_V V)^{\ba} 
	= \Big[ G^{-1} \cdot \big[ L_V (G\cdot V) - \tfrac12 \dd [G(V,V)] \big] \Big]^{\ba} 
	= 2 V^{a}  D_{a} V^{\ba}
\end{equation}
Thus we see that equations~\eqref{eq:gen-geo1} are the analogue of the auto-parallel condition, given by~\eqref{eq:gen-auto-parallel}. 

Let us also note, that we can recover~\eqref{eq:string-eom-vtv} directly, without explicit use of generalised connections. 
As $E \simeq E^*$, the generalised covector $G\cdot V$ can also be viewed as a generalised vector, with untwisted picture components
\begin{equation}
	\eta^{-1} \cdot G \cdot V = \tv + g\cdot v
\end{equation}
We thus have that
\begin{equation}
\begin{aligned}
	L_V (\eta^{-1} \cdot G\cdot V) &= [v,\tv] + \mathcal{L}_v (g\cdot v) - i_{\tv} \dd(g\cdot \tv) - i_v i_{\tv} H \\
	&= [v,\tv] + \Big[ v^p \LC_p v_m + (\LC_m v^p) v_p - \tv^p(\LC_p \tv_m - \LC_m \tv_p)
		+ H_{mnp} v^n \tv^p \Big] \dd x^m \\
	&= [v,\tv] + \Big[ v^p \LC_p v_m - \tv^p\LC_p \tv_m 
		+ H_{mnp} v^n \tv^p  \Big] \dd x^m
		+ \tfrac12 \der_m (v^p v_p  + \tv^p \tv_p) \dd x^m
\end{aligned}
\end{equation}
Again, with careful consideration of the numerical factors, we find
\begin{equation}
\begin{aligned}
	\eta^{-1} \cdot \Big[ L_V (G\cdot V) - \tfrac12 \dd G(V,V) \Big]
	= [v,\tv] + \Big[ v^p \LC_p v_m - \tv^p\LC_p \tv_m 
		+ H_{mnp} v^n \tv^p  \Big] \dd x^m = 0
\end{aligned}
\end{equation}
becomes the equation of motion~\eqref{eq:string-eom-vtv} together with the vanishing of the commutator $[v,\tv]$.


\subsection{Physical interpretation}
\label{sec:phys}

To understand why~\eqref{eq:new-auto-parallel} is a physically natural formulation of the auto-parallel condition, we consider the action of a particle in general relativity:
\begin{equation}
	S = \int \dd \lambda \left( 
		\frac12 g_{mn} \frac{\der x^m}{\der \lambda} \frac{\der x^n}{\der \lambda} \right)
\end{equation}
Writing $v^m = \frac{\der x^m}{\der \lambda}$ and imagining that we have a congruence of trajectories as above, the conjugate momentum to the coordinate $x^m$ and the Hamiltonian are thus
\begin{equation}
\begin{aligned}
	p_m = g_{mn} v^m \qquad 
		\mathcal{H} = p_m v^m - L = \tfrac12 g_{mn} v^m v^n 
			= \tfrac12 g(v,v) = \tfrac12 g^{-1}(p,p)
\end{aligned}
\end{equation}
so that the auto-parallel condition, formulated as~\eqref{eq:new-auto-parallel}, is the statement that
\begin{equation}
\label{eq:Lvp}
	\mathcal{L}_v p = \dd \mathcal{H}
\end{equation}
This equation is reminiscent of one of Hamiltons equations (usually written in textbooks as $\dot{p} = - \frac{\der \mathcal{H}}{\der q}$) but with the simple time derivative replaced by the Lie derivative along the flow generated by the vector field $v$. 
This is natural, as the vector field $v$ defines a Hamiltonian ``time" coordinate $\lambda$ on the target spacetime. If we work in a coordinate system with $\lambda$ as the first coordinate, in which $v = \frac{\der}{\der \lambda}$, then the Lie derivative expression reduces to the partial derivative with respect to this coordinate on the components of tensor fields. The Lie derivative thus provides a covariantisation of the derivative with respect to Hamiltonian ``time" $\lambda$. 
Put another way, the Lie derivative is a (covariant) time derivative which is natural if we match the gauge for spacetime diffeomorphisms to the gauge for worldline diffeomorphisms. This is often referred to as static gauge in the literature. 

However, the interpretation is slightly different to the usual Hamiltonian formalism in other ways. The Hamiltonian is usually thought of as a function on the phase space of the system, expressed in terms of the coordinates and conjugate momenta. In our case, the Hamiltonian depends on the coordinates only through the inverse metric, and is quadratic in the momenta. However, in~\eqref{eq:Lvp}, the Hamiltonian is thought of simply as a function on the spacetime manifold, albeit a function which is expressed in terms of the value of the vector field $v$. One could thus also compare this to a simple potential force law. The left side could be thought of as the force (rate of change of momentum), while the right side is the gradient of the potential energy for that force. In this case the potential and the momentum are both written in terms of the same vector field $v$, together with the metric.

We also note that given a Killing vector $k$, the usual statement of the conservation law can also be seen easily from~\eqref{eq:Lvp}, without introducing a connection. We have:
\begin{equation}
\begin{aligned}
	\frac{\der}{\der \lambda} \big[ g(v,k) \big] &= \mathcal{L}_v \langle p , k \rangle \\
		&= \langle \dd \mathcal{H} , k \rangle + \langle p , \mathcal{L}_v k \rangle \\
		&= \mathcal{L}_k \mathcal{H} + \langle p , \mathcal{L}_v k \rangle \\
		&= g( v, \mathcal{L}_k v) + \langle p , \mathcal{L}_v k \rangle \\
		&= \langle p , \mathcal{L}_k v + \mathcal{L}_v k \rangle \\
		&= 0
\end{aligned}
\end{equation}

We will now show that all of this works in the same way for the string, but with a generalised notion of momentum. First, we examine the conjugate momentum to the coordinate $x^m$ for the string action~\eqref{eq:string-action}. This is given by
\begin{equation}
	\rho_m = g_{mn} v^m - B_{mn} \tv^n 
\end{equation}
Here we immediately encounter an apparent difference from the situation above: this quantity is not gauge-invariant with respect to $B$-field gauge transformations. This is typical of conjugate momenta in the presence of gauge fields, and one encounters the same for a particle coupled to an electromagnetic vector potential $A_m$. As we change gauge via $B' = B + \dd \Lambda$, we have that $\rho$ transforms as 
\begin{equation}
	\rho' = \rho - i_{\tv} \dd \Lambda
\end{equation}
and thus, the one-form $\rho$ transforms as the one-form part of a generalised vector in the twisted picture. The vector part of this generalised vector is $\tv$, the tangent vector in the spacelike direction along the string. We can think of this as being a momentum dual to the ``winding" or charge of the string, though here we do not assume any circle directions or isometries in the spacetime. This motivates the definition of the momentum generalised vector (in the twisted picture) as:
\begin{equation}
	P = \tv + \rho = \tv + g \cdot v - i_{\tv} B
\end{equation}
(This object was previously identified in the discussion of~\cite{Blair:2013noa}.) If we move to the untwisted picture representation this becomes simply
\begin{equation}
\label{eq:P-E-def}
	P = \tv + \rho = \tv + g \cdot v
\end{equation}
which is the generalised vector $V$ with $v$ and $\tv$ interchanged. This can be written in $O(10,10)$ indices as 
\begin{equation}
	P^M = \eta^{MN} G_{NP} V^P
\end{equation}
so that identifying $E \simeq E^*$ (i.e. using the $O(10,10)$ metric to raise and lower indices) we have the generalised covector 
\begin{equation}
	P = G \cdot V
\end{equation}
Noting that the Hamiltonian of the string is
\begin{equation}
		\mathcal{H} = \rho_m v^m - L = \tfrac12 g_{mn} ( v^m v^n + \tv^m \tv^n)
			=  G(V,V)
			= G^{-1} (P, P)
\end{equation}
we now see that the auto-parallel condition for the generalised tangent vector $V$ for the string from~\eqref{eq:gen-auto-parallel} becomes (viewing $\dd \cH$ as a section of $E^*$ as above~\eqref{eq:dG-Estar})\footnote{The unsightly factor of $1/2$ in this equation stems from the normalisation of the $O(10,10)$ metric $\eta$. It is removed on viewing $P$ and $\dd \cH$ instead as sections of $E$ as below.}
\begin{equation}
\label{eq:Odd-LVP}
	L_V P = \tfrac12 \dd \mathcal{H}
\end{equation}
which is clearly the precise analogue of equation~\eqref{eq:Lvp} for the generalised geometry system. Thus, we have the statement that the infinitesimal flow of the generalised momentum along the generalised diffeomorphism generated by $V$ is equal to the gradient of the Hamiltonian. 
If we fix the gauge for spacetime diffeomorphisms, i.e. our local coordinates, to be such that the worldsheet coordinates $\tau$ and $\sigma$ are the first two coordinates, thus matching the gauge on the worldsheet, then one could expect that the Dorfman derivative will involve only simple partial derivatives along those directions. In fact this is only true if one also imposes the Virasoro constraint $g(v,\tv)=0$. 

We can also view $P$ naturally as a section of $E$ as we originally defined it in~\eqref{eq:P-E-def}, and also $\dd \cH$ as a section of $E$ with normalisation $\eta(V, \dd\cH) = \tfrac12 i_v \dd\cH$. Doing so,~\eqref{eq:Odd-LVP} becomes
\begin{equation}
\label{eq:Odd-LVP2}
	L_V P = \dd \mathcal{H}
\end{equation}
From this, we also see that, just as the equations of motion are symmetric in the interchange of the worldsheet coordinates $\tau$ and $\sigma$, this equation is symmetric in $V$ interchanged with $P$. In particular, 
\begin{equation}
	L_P V = 2\dd \eta(P,V) - L_V P = 2\dd G(V,V) -  \dd G(V,V) = \dd G(V,V) 
		= \dd \mathcal{H}
\end{equation}
so that in fact the equation can be written as the vanishing of the Courant bracket (i.e. the anti-symmetric part of the Dorfman derivative)
\begin{equation}
	[V, P ]  = 0
\end{equation}

Naively, the exchange of $\sigma$ and $\tau$ appears to resemble the ingredients of a T-duality transformation. However, this should not be confused with T-duality, as we do not change the background metric and $B$-field. It is simply the exchange of $\tau$ and $\sigma$ on the worldsheet. 

Let us also examine the possible analogue of the conservation law for a generalised Killing vector $K$ with $L_K G = 0$~\cite{Grana:2008yw,CS2}. We have
\begin{equation}
\label{eq:gen-conservation}
\begin{aligned}
	\frac{\der}{\der \tau} \big[ G(V,K) \big] &= L_V \langle P , K \rangle \\
		&= \langle \dd \mathcal{H} , K \rangle + \langle P ,L_V K \rangle \\
		&= \tfrac12 L_K \mathcal{H} + \langle P , L_V K \rangle \\
		&= G( V, L_K V) + \langle P , L_V K \rangle \\
		&= \langle P , L_K V + L_V K \rangle \\
		&= 2\langle P , \dd \eta(K,V) \rangle \\
		&= L_P \langle V , K \rangle \\
		&= \frac{\der}{\der \sigma} \langle V , K \rangle
\end{aligned}
\end{equation}
Up until the last steps, this is identical to the calculation for the ordinary geodesic above, but unlike the Lie bracket, the Dorfman derivative is not anti-symmetric. This reflects that the generalised tangent space is in general a Leibnitz algebroid~\cite{Baraglia}, rather than a Lie algebroid, and is related to 
the tensor hierarchy of gauge transformations of the $B$-field~\cite{Hohm:2013nja, deWit:2005hv, deWit:2008ta,Lavau:2017tvi,Kotov:2018vcz} and its associated $L_\infty$ structure~\cite{Hohm:2017pnh, RoytenbergWeinstein, Jurco:2018sby, Cederwall:2018aab,  Bonezzi:2019ygf, Lavau:2020pwa}. We thus do not get automatic conservation of the inner product of $K$ with the generalised tangent vector $V$, but we do have conservation if $\eta(V,K) = 0$ (for example if $K \propto V$ given that $V$ satisfies the Virasoro constraints).

Finally, let us note that, similarly to in ordinary geometry where a null Killing vector is automatically auto-parallel, a generalised Killing vector satisfying $\eta(K,K) = G(K,K) = 0$ is also auto-parallel in our generalised sense. 
Thus, such generalised Killing vectors may give rise to string worldsheet solutions. As we discuss in the conclusion, whether we get a string or not depends on whether both the vector and one-form components of $K$ are non-vanishing. The condition for this is simply that $g(k,k) \neq 0$ where $k$ is the vector component of $K$.


\section{M2 branes restricted to four-dimensional space}
\label{sec:M2}

In this section we provide the corresponding construction for the M2 brane. We find that the exact same conditions on a generalised vector field in a different version of generalised geometry, with generalised structure group $\SL(5,\bbR)\times\bbR^+$, reproduce the equations of motion for a gauge-fixed formulation of the worldvolume of the M2 brane, restricted to a four-dimensional sector of a dimensional split (which is necessary for the formulation of the exceptional generalised geometry). In contrast to most constructions of exceptional generalised geometry, we include the time direction in this four-dimensional sector, such that the generalised metric defines an $\SO(3,2)$ subgroup of $\SL(5,\bbR)\times\bbR^+$. 


\subsection{The gauge-fixed M2 brane theory}

The bosonic part of the M2 brane action is~\cite{Bergshoeff:1987qx}
\begin{equation}
\begin{aligned}
	S = \int \dd^3 \sigma \Big[ -\tfrac12 \sqrt{-\gamma} \Big( \gamma^{\alpha\beta} g_{mn} \der_\alpha x^m \der_\beta x^n 
	- 1 \Big)
	+ \tfrac{1}{3!} \epsilon^{\alpha\beta\gamma} A_{mnp} \der_\alpha x^m \der_\beta x^n \der_\gamma x^p
	\Big]
\end{aligned}
\end{equation}
which gives rise to the equations of motion for the field $x^m$
\begin{equation}
\label{eq:M2-x-eom}
	\LC_\alpha \der^\alpha x^m 
		+ \Gamma_p{}^m{}_q \gamma^{\alpha \beta} \der_\alpha x^p \der_\beta x^q
		- \tfrac{1}{3!} \epsilon^{\alpha \beta\gamma} F^m{}_{npq}  
			\der_\alpha x^n \der_\beta x^p \der_\gamma x^q = 0 
\end{equation}
where $\LC_\alpha$ is the worldvolume Levi-Civita connection, and 
\begin{equation}
\label{eq:M2-gamma-eom}
	\gamma_{\alpha\beta} = g_{mn} \der_\alpha x^m \der_\beta x^n
\end{equation}
for the worldvolume metric. Note that the equation of motion for the worldvolume metric sets it equal to the pullback of the target spacetime metric to the worldvolume. This complicates the theory substantially. 

Here, we use the formulation of the M2 brane in which the worldvolume diffeomorphisms are gauge-fixed (as in~\cite{Hartnoll:2002th}) so that the metric $\gamma_{\alpha\beta}$ has 
\begin{equation}
\label{eq:M2-gauge}
	\gamma_{0i} = 0 \hs{30pt}  \gamma_{00} + \det [\gamma_{ij}] = 0
\end{equation}
for indices $\alpha = (0,i)$ and $i=1,2$. This results in the action
\begin{equation}
\begin{aligned}
	S = \int \dd^3 \sigma \Big[ \tfrac12 \Big( g_{mn} \der_0 x^m \der_0 x^n 
	- \tfrac12 g_{mn} g_{m'n'} \lambda^{mm'} \lambda^{nn'} \Big)
	+ \tfrac{1}{3!} \epsilon^{\alpha\beta\gamma} A_{mnp} \der_\alpha x^m \der_\beta x^n \der_\gamma x^p
	\Big]
\end{aligned}
\end{equation}
where, anticipating what is to come, we use the shorthand $\lambda^{mn} = \epsilon^{ij} \der_i x^m \der_j x^m$ with the convention that $\epsilon^{12} = +1$.

The conjugate momentum for the coordinate $x^m$ is then the target space one-form\footnote{This is derived using the convention that on the Lorentzian worldsheet $\epsilon_{012} = +1$ and $\epsilon^{012} = -1$.}
\begin{equation}
\label{eq:M2-momentum}
	p_m = g_{mn} \der_0 x^n - \tfrac12 A_{mnp} \lambda^{np}
\end{equation}
which, as for the string, is not gauge invariant under $A' = A + \dd \Lambda$. The Hamiltonian is then given by 
\begin{equation}
\label{eq:M2-Ham}
	\cH = \tfrac12 g(v_0, v_0) + \tfrac14 g_{mn} g_{m'n'} \lambda^{mm'} \lambda^{nn'}
\end{equation}
where we have written $v_0^m = \der_0 x^m$. The equation of motion for the field $x^m$ takes the form
\begin{equation}
\label{eq:M2-eom}
	v_0^p \LC_p v_{0m} 
		+ \lambda^{np} \LC_n \lambda_{pm} 
		+ \tfrac12 F_{mnpq} v_0^n \lambda^{pq} = 0
\end{equation}
where we have extended $v_0$ and $\lambda$ to fields on a patch of spacetime, as for equation~\eqref{eq:string-eom-vtv}. 


\subsection{$\SL(5,\bbR) \times \bbR^+$ generalised geometry with Lorentz group $\SO(3,2)$}

We will work in a spacetime with a metric of the form
\begin{equation}
	\dd s^2 = \ee^{2\Delta} \delta_{\mu\nu} \dd y^\mu \dd y^\nu + g_{mn} \dd x^m \dd x^n
\end{equation}
with $m,n = 0,1,2,3$ the indices on the ``internal" part of the space. We also take the three-form field $A_{(3)}$ to have components $A_{mnp}$ along the internal directions, and impose that all fields depend only on the internal coordinates $x^m$. This is a typical ansatz for a warped compactification to seven-dimensional Minkowski space, but here we take the internal metric $g_{mn}$ to have Lorentzian signature, while the external factor is warped Euclidean. For simplicity, in this paper we will set the warp factor to zero. 

The theory of eleven-dimensional supergravity restricted to the four-dimensional ansatz outlined above admits a description in terms of a generalised geometry~\cite{chris,PW,CSW2,BermanPerry} with structure group $E_{4(4)} \simeq \SL(5,\bbR)\times\bbR^+$ and generalised tangent space
\begin{equation}
	E \simeq T \oplus \Lambda^2 T^*
\end{equation}
A generalised vector field $V \in \Gamma(E)$ is given by a vector field $v$ together with a collection of two-forms which transform under the gauge transformations $A' = A + \dd \lambda$ via 
\begin{equation}
\label{eq:SL5-gauge}
	\lambda' = \lambda + i_v \dd \Lambda
\end{equation}
The key objects which we need here are the Dorfman derivative (or generalised Lie derivative) which for generalised vectors $V = v + \lambda$ and $V' = v' + \lambda'$ is given by
\begin{equation}
	L_V V' = \mathcal{L}_v v' + (\mathcal{L}_v \lambda' - i_{v'} \dd \lambda)
\end{equation}
and the generalised metric
\begin{equation}
	G(V,V) = (g_{mn} + \tfrac12 A^{pq}{}_m A_{pqn}) v^m v^n - A^{pq}{}_m v^m \lambda_{pq} + \tfrac12 \lambda^{pq} \lambda_{pq}
\end{equation}
where indices are raised and lowered with the ordinary metric $g_{mn}$. As here we take $g_{mn}$ to be Lorentzian, the generalised metric here is stabilised by $\SO(3,2) \subset \SL(5,\bbR)$. 

We will also be interested in sections $W = \zeta + \beta$ of $E^* \simeq T^* \oplus \Lambda^2 T$. The one-form parts $\zeta$ of these transform under~\eqref{eq:SL5-gauge} as 
\begin{equation}
\label{eq:SL5-gauge-Estar}
	\zeta' = \zeta - \beta \inn \dd \Lambda
\end{equation}
which can be verified as this leaves the natural inner product\footnote{The symbol $\inn$ denotes the contraction of a multivector into a form with the same conventions as in~\cite{CSW2}.}
\begin{equation}
	\langle W, V \rangle = v \inn \zeta + \beta \inn \lambda
\end{equation}
invariant. 

As for the discussion of $O(10,10)$ generalised geometry in section~\ref{sec:gen-geom-Odd}, 
there are two descriptions of generalised vectors which are equally good. The description above, with two-forms transforming under gauge transformations, is the twisted picture of this geometry. 
Given such a generalised vector $V = v + \lambda$, we can also define
\begin{equation}
	V_{\text{Untwisted}} = v + \lambda - i_v A
\end{equation}
which is a global section of $T \oplus \Lambda^2 T^*$, thus realising the isomorphism $E \simeq T \oplus \Lambda^2 T^*$. In~\cite{CSW2}, this is described in terms of the components with respect to particular frames called split frames. In the untwisted picture, the generalised metric and Dorfman derivative are given by different explicit formulae to those above. We have instead 
\begin{equation}
	L_V V' = \mathcal{L}_v v' + (\mathcal{L}_v \lambda' - i_{v'} \dd \lambda) -  i_v i_{v'} F
\end{equation}
and the generalised metric becomes simply
\begin{equation}
	G(V,V) = g_{mn} v^m v^n + \tfrac12 \lambda^{pq} \lambda_{pq}
\end{equation}
In what follows, we will also be interested in an expression for the Dorfman derivative acting on a section of $E^*$. Written in terms of the untwisted objects $V = v + \lambda$ and $W = \zeta + \beta$, this is given by 
\begin{equation}
\label{eq:LVEstar}
	L_V W = \Big[ \mathcal{L}_v \zeta + \beta \inn \dd \lambda - v \inn (\beta \inn F) \Big]
		+ \mathcal{L}_v \beta
\end{equation}

There is also another bundle, denoted by $N$, which contains the parameters for the gauge-transformations of the gauge-transformations in the supergravity~\cite{CSW2}. For the $\SL(5,\bbR)\times\bbR^+$ generalised geometry, its fibre transforms in the $\rep{5'_{+2}}$ representation of $\SL(5,\bbR)\times\bbR^+$~\cite{Berman:2011cg}, where the generalised tangent space transforms in the $\rep{10_{+1}}$. The formula,~\eqref{eq:Dorfman-sym} for the symmetric part of the Dorfman derivative generalises to the statement that 
\begin{equation}
	L_V V' + L_{V'} V = \der \proj{E} (V \proj{N} V')
\end{equation}
where the symbol $\proj{X}$ means that one takes the tensor product and then projects (covariantly) onto the part in the bundle $X$.\footnote{For $O(d,d)$ generalised geometry, $N$ is the trivial real line bundle so that $V \proj{N} V'$ is simply the scalar function $2\langle V, V' \rangle$.} 

In constructions of extended geometry, where one looks to enlarge the spacetime by adding additional coordinates corresponding to the non-vector directions in the generalised tangent space, the physical spacetimes (or duality frames) are often defined by restricting the dependence of fields to a set of  directions which are mutually null in the section condition. In other words, the derivatives satisfy $\der X \proj{N^*} \der Y = 0$ for all fields and parameters $X$ and $Y$. This is referred to as the strong section condition, and it is needed for the closure of the algebra of generalised diffeomorphisms in these constructions~\cite{Berman:2012vc}.  


\subsection{Equations of motion from generalised geodesics}

We claim that the same generalised geodesic equation as before, 
\begin{equation}
\label{eq:LVP-SL5}
	L_V P = \dd \cH
\end{equation}
encodes the equations of motion for integral surfaces of the generalised vector $V$ which match the gauge-fixed M2 brane theory.

To make this claim, first we need to understand how the generalised vector encodes three directions in the four-dimensional space. To see this, let us impose the condition that $V$ is null in the projection to $N$, i.e.
\begin{equation}
	V \proj{N} V = 0
\end{equation}
This is the analogue of the Virasoro condition $\eta(V,V) = 0$ for the string. In terms of the vector and two-form components of $V$ in the untwisted picture, it says that 
\begin{equation}
\label{eq:4d-section-cpts}
	i_v \lambda = 0
	\hs{30pt}
	\lambda \wedge \lambda = 0
\end{equation}
The second condition implies that $\lambda$ is rank one and thus $\lambda = \lambda_1 \wedge \lambda_2$ for two one-forms $\lambda_1$ and $\lambda_2$. Via the four dimensional metric these give rise to two vectors $v_1 = g^{-1}(\lambda_1, \cdot)$ and $v_2 = g^{-1}(\lambda_2, \cdot)$. 
These vectors must be linearly independent for a non-zero two-form $\lambda$. 
Further, the other condition in~\eqref{eq:4d-section-cpts} says that the vector $v$ is orthogonal to $v_1$ and $v_2$ in the metric, matching the first gauge fixing condition for the worldvolume metric~\eqref{eq:M2-gauge}. 
Assuming that $v$ is timelike, $v_1$ and $v_2$ are then spacelike and, without loss of generality, orthogonal. 
Thus, the generalised vector satisfying $V \proj{N} V = 0$ becomes equivalent to three vectors which are orthogonal to each other in the four-dimensional metric. These will become the three directions along the worldvolume of the M2 brane. 
Imposing also that $V$ is null in the generalised metric $G(V,V) = 0$ as for the string, we find that 
\begin{equation}
	g_{mn} v^m v^n + \tfrac12 g^{mn}g^{m'n'} \lambda_{mm'} \lambda_{nn'} = 0
\end{equation}
which matches the second gauge condition for the worldvolume metric~\eqref{eq:M2-gauge}. 
This is solved if $v$ is timelike and $v_1$ and $v_2$ are spacelike with appropriately related magnitudes. As for the string, we assume that the components of our generalised vector fit this pattern and are labelled ``non-degenerate" as such. Though there are other configurations which would solve the same constraints, and we cannot impose this type of non-degeneracy at the level of $\SL(5,\bbR)$ covariant conditions, we leave this issue for the discussion.

Next, we must examine the conjugate momentum as above. As for the string, the natural one-form~\eqref{eq:M2-momentum} has the correct gauge transformation under $A' = A + \dd \Lambda$ to be the one-form component of a local section of $E^* \simeq T^* \oplus \Lambda^2 T$ in the twisted picture. In the untwisted picture, this section is given by the global relation
\begin{equation}
\label{eq:SL5-P}
	P = g \cdot v + v_1 \wedge v_2 = G \cdot V
\end{equation}
In terms of these objects, it is clear that the Hamiltonian~\eqref{eq:M2-Ham} takes the form
\begin{equation}
\label{eq:SL5-H}
	\cH = \tfrac12 G(V,V) = \tfrac12 G^{-1} (P,P)
\end{equation}
However, note that the symmetry between $V$ and $P$ that we found in section~\ref{sec:phys} is special to the case of the string and does not have any analogue here.
As for our discussion of the string, similar objects to~\eqref{eq:SL5-P} and~\eqref{eq:SL5-H} have appeared in previous works looking at membrane sigma models in extended geometries~\cite{Hatsuda:2012vm,Sakatani:2016sko,Sakatani:2017vbd}. 

It thus remains to show that~\eqref{eq:LVP-SL5} indeed encapsulates the equations of motion. Using~\eqref{eq:LVEstar} we have that as a section of $E^*$, in the untwisted picture:
\begin{equation}
\begin{aligned}[]
	[L_V (G \cdot V)]_m &= v^p (\LC_p v_m + \LC_m v^p) + \tfrac12 \lambda^{pq} (\dd \lambda)_{pqm} 
		+ \tfrac12 F_{mnpq} v^n \lambda^{pq} \\
	[L_V (G \cdot V)]^{mn} &= (\mathcal{L}_v \beta)^{mn} 
\end{aligned}
\end{equation}
where $\beta^{mn} = g^{mp} g^{nq} \lambda_{pq}$ are the components of the bivector part of $G \cdot V$. 
Equation~\eqref{eq:LVP-SL5} then becomes
\begin{equation}
\begin{aligned}[]
	v^p (\LC_p v_m + \LC_m v^p) + \tfrac12 \lambda^{pq} (\dd \lambda)_{pqm} 
		+ \tfrac12 F_{mnpq} v^n \lambda^{pq} 
		&= \frac12 \LC_m \big(v^p v_p + \tfrac12 \lambda^{pq} \lambda_{pq}\big) \\
	\mathcal{L}_v \beta &= 0
\end{aligned}
\end{equation}
Via some simple manipulations, the first of these equations becomes equivalent to~\eqref{eq:M2-eom}. Setting $v^m = v^m_0$ and $\beta^{mn} = 2v_1^{[m} v_2^{n]}$ as suggested above, the second equation becomes
\begin{equation}
	[v_0, v_1] \wedge v_{2} + v_{1}\wedge  [v_0, v_2]  = 0
\end{equation}
which is solved as $[v_0,v_1] = [v_0, v_2] = 0$ if the vectors $v_0, v_1, v_2$ can be assigned coordinates, as we require for our worldvolume. This completes the demonstration that~\eqref{eq:LVP-SL5} encapsulates the equations of motion of the M2 brane in this formulation of generalised geometry.

We also briefly note that the manipulations~\eqref{eq:gen-conservation} in this case lead to 
\begin{equation}
\label{eq:gen-conservation-M2}
\begin{aligned}
	\frac{\der}{\der \tau} \big[ G(V,K) \big] &= L_V \langle P , K \rangle \\
		&= \langle P , \der \proj{E} (K \proj{N} V) \rangle \\
\end{aligned}
\end{equation}
so that we also have a conservation law here, provided that our generalised Killing vector $K$ has $K \proj{N} V = 0$. Any generalised Killing vector which is null in the generalised metric and has $K \proj{N} K = 0$ will automatically be auto-parallel in this case too. 

Also, while equation~\eqref{eq:LVP-SL5} naively appears to involve derivatives in directions other than those along the worldvolume, when expanded we find that this is not the case. 
This could be made manifest by expanding the equation in terms of a generalised Levi-Civita connection~\cite{CSW2} for the $\SO(3,2)$ generalised metric, as we did in section~\ref{sec:gen-string-2} for the string.


\section{Discussion}
\label{sec:discussion}

A summary of our main result is as follows. The classical worldvolume of a string or brane in a background generalised metric $G$ has an associated generalised tangent vector $V$ which is null in the generalised metric and section condition
\begin{equation}
\label{eq:null}
	G(V,V) = 0
	\hspace{30pt}
	V \proj{N} V = 0
\end{equation}
When extended to a generalised vector field on an open set containing the worldvolume, it solves the equation of motion
\begin{equation}
\label{eq:gen-eom}
	L_V P = \dd \mathcal{H}
\end{equation}
on the worldvolume, where $P = G \cdot V$ is the generalised conjugate momentum covector and $\mathcal{H} = \tfrac12 G(V,V) = \tfrac12 G^{-1} (P,P)$ is the Hamiltonian/energy function on the spacetime. 
Equation~\eqref{eq:gen-eom} can be thought of as the analogue of the auto-parallel condition for the generalised vector field $V$. 
The extension of $V$ to a local vector field is technically necessary for the Dorfman derivative to be defined, though the equations need only be solved on the worldvolume. If the equations are solved everywhere, then we obtain a foliation of spacetime by worldvolume solutions, at least locally.

In the case of the string, we can argue a converse result: given a generalised vector field $V$ satisfying these equations which is non-singular in an appropriate sense to be discussed, one obtains a foliation of spacetime by classical worldsheet solutions. We have not firmly established this converse result for the membrane, though one could expect that it will be confirmed by further analysis. 

This converse follows from considering the solutions to equations~\eqref{eq:null} and~\eqref{eq:gen-eom} for a generalised vector field $V$. Let us suppose that both vectors $v$ and $\tv$ derived from the generalised vector are non-vanishing on some local patch of spacetime. Since $[v,\tv] = 0$, they define a folitation into two-dimensional sheets, and we can choose coordinates on the sheets such that they are coordinate induced vectors. Since $\eta(V,V) = 0 \Ra g(v,\tv) = 0$, they are orthogonal and $G(V,V) = g(v,v) + g(\tv, \tv) = 0$ implies that they are either both null or one is spacelike and the other timelike. If they are both null, then orthogonality implies that they are proportional, and thus the worldsheet degenerates to be one-dimensional. 
Assuming that this does not happen, i.e. that $g(v,v) \neq 0$, 
we thus recover that the worldsheet coordinates $\tau$ and $\sigma$ are the coordinates inducing $v$ and $\tv$. 

Throughout the paper, we have avoided the question of how, in addition to the constraints~\eqref{eq:null}, one could specify that the generalised vector field associated to the foliation is non-singular in this way, and so gives a two-dimensional sheet in the string case. 
In fact, this question seems to be rich with possibility. Consider, for example, the type IIA decomposition of $\SL(5,\bbR)\times\bbR^+$ generalised geometry (see~\cite{CdFPSW} for a full presentation of type IIA decompositions). There the generalised tangent space decomposes as
\begin{equation}
\begin{aligned}
	E &\simeq T \oplus T^* \oplus \bbR \oplus \Lambda^2 T^* \\
	V &= v + \lambda + \omega_0 + \omega_2
\end{aligned}
\end{equation}
and the condition $V \proj{N} V = 0$ becomes
\begin{equation}
	i_v \lambda = 0
	\hs{30pt}
	\omega_2 \wedge \omega_2 = 0
	\hs{30pt}
	i_v \omega_2 + \omega_0 \lambda = 0
\end{equation}
while $G(V,V) = 0$ becomes
\begin{equation}
	G(V,V) = v^p v_p + \lambda^p \lambda_p + (\omega_0)^2 
		+ \tfrac12 (\omega_2)^{pq} (\omega_2)_{pq} = 0
\end{equation}
This gives rise to many possibilities. For example, we could have only $v$ and $\lambda$ non-zero and be left with a system like our treatment of the string above (but including background RR fluxes). Alternatively, we could have only $v$ and $\omega_2$ non-zero, which would give a system for the D2 brane, similar to our picture for the M2 brane above. As the D2 (and the type IIA decomposition of exceptional geometry) is the straightforward dimensional reduction of the unwrapped M2 system, the equation of motion should match. Another possibility would have only $v$ and $\omega_0$, giving the D0 brane. The case where only $v$ itself is non-zero (and thus a null vector) would presumably correspond to an infinite boost limit of any of these objects, and thus a pure momentum state. This last statement seems universal. 

One could thus hope that these conditions will give all possible objects described by the relevant duality group, including objects such as the M5 and NS5 branes for the higher rank exceptional groups, and that~\eqref{eq:gen-eom} will provide all of their equations of motion. 
One potential complication could be the inclusion of the gauge fields that appear on the worldvolumes of these objects. However, there is reason to be hopeful there also: the generalised tangent vectors in those cases include more degrees of freedom than purely the directions along the worldvolumes. 
For example, if one considers D-branes in type II theories, the generalised vector contains a one-form which could encode the one-form gauge field. Further, for the type IIB NS5 brane there is an additional one-form and for the M5 and type IIA NS5 there is a two-form. Thus, the generalised vectors in question appear to have the degrees of freedom to include the relevant gauge fields. 
The corresponding charges for these components of the generalised vectors correspond to the objects which end on the relevant brane, e.g. the string charge in the case of D-branes and the M2 charge in the case of M5 branes, as one would expect. 

Note that our description has advantages over previous formulations of membrane sigma models~\cite{Hatsuda:2012vm,Sakatani:2016sko,Sakatani:2017vbd} in this respect. Dualities change the dimensions of the worldvolumes of branes. For example, T duality on a circle shifts the dimension of a D-brane up or down by one depending on whether the worldvolume wraps the circle or not. 
This makes it problematic to formulate a ``duality invariant" action, if the dimension of the worldvolume one integrates over is fixed.
However, our generalised vector field description does not suffer from this problem, as the dimension of the brane is determined by the decomposition of the generalised vector in the duality frame in question. 
One could also hope that the equations could include the possibility of type-changing solutions, which degenerate along different loci and contain components with different dimensions, at least in limiting cases.  
This might then describe intersections of different objects, such as strings ending on branes, all as the solution to a single set of equations. 

We should note that some of the above observations concerning the possible solutions of the algebraic equations~\eqref{eq:null} are fairly well known in the literature (see~\cite{Obers:1998fb,Berman:2020tqn} and references therein): they can be seen as the BPS condition and the $\tfrac12$-BPS condition. 
There is also a clear link here to the recent works~\cite{Berkeley:2014nza,Berman:2014jsa,Berman:2014hna,Berman:2020tqn} in which the target space supergravity solutions of flat branes are seen to correspond to plane waves in some duality frame. The condition $V \proj{N} V = 0$ is usually referred to as the section condition in extended geometries, and subspaces of vectors which mutually satisfy it correspond to the spacetime directions of the various duality frames (sometimes also called polarisations) in the extended space. In~\cite{Berkeley:2014nza,Berman:2014jsa,Berman:2014hna,Berman:2020tqn} it is shown that the different duality frame perspectives on a null wave in the extended spacetime give the supergravity solutions corresponding to the various branes whose charges are included in the generalised tangent space. The duality frame in which the solution corresponds to a wave is one in which the generalised vector $V$ is of the pure vector type (and thus manifestly satisfies $V \proj{N} V = 0$).

Our result here could be thought of as a generalisation of this statement from the perspective of the worldvolume theories and within the realm of generalised geometry, where the anchor map $\pi : E \ra T$ fixes the duality frame. However, whereas the solutions of~\cite{Berkeley:2014nza,Berman:2014jsa,Berman:2014hna,Berman:2020tqn} describe only the flat worldvolumes corresponding to the standard supersymmetric brane solutions in flat space, our result contains arbitrary solutions of the worldvolume theories in arbitrary backgrounds (restricted to the dimensions included in the generalised geometry in the exceptional case). 
For a generalised vector with only a null vector component, our equations describe ordinary null geodesics, corresponding to null waves. In a sense, our result could be viewed simply as writing that system in generalised geometry covariant language. Having done this, it then seems inevitable by symmetry that the other solutions to~\eqref{eq:null} will correspond to the objects carrying the other charges. 

It would also be interesting to link our results with supersymmetry. Supersymmetric branes are encoded in the spacetime geometry (with fluxes) by (generalised) calibrations~\cite{Gualtieri,Harvey:1982xk,Becker:1995kb, Gutowski:1999tu, Martucci:2005ht}. In~\cite{deFelice:2017mhm}, it was shown that the calibration forms are related to the pull-backs to the branes of the generalised Killing vectors which arise as the commutators of supersymmetries on the backgrounds (see~\cite{CS2,CS3} for details of these superalgebras in generalised geometry language), thus confirming a conjecture from~\cite{APW}. Thus, in our language it could be that these supersymmetric worldvolumes correspond precisely to those generalised Killing vectors. 

Another point which goes beyond the scope of the present work concerns global aspects of our worldvolumes. We have worked entirely within a gauge-fixed framework i.e. some particular choice of coordinates on the worldvolume, which may not be available globally. Though our expressions~\eqref{eq:null} and~\eqref{eq:gen-eom} are manifestly coordinate-free on the target space, we must wonder how to patch together our generalised vectors on the overlaps of these patches of the worldvolume. We leave this issue for future consideration.

An intriguing but speculative possible extension of these ideas concerns ``higher" geometry. While an ordinary vector defines a one-parameter family of diffeomorphisms along which the particles flow as they propagate, the generalised vector is more complicated. For example, in the string case, it encodes a diffeomorphism and a gauge transformation of the two-form field $B$ which together are a general bosonic symmetry of the background fields. However, $V$ should be thought of as an element of an $L_3$ algebra rather than a Lie algebra. Thus, it is not clear what a generalised vector ``integrates to". Previous efforts to exponentiate generalised Lie derivatives in the context of Double Field Theory~\cite{Hohm:2012gk} have led authors to consider exotic forms of geometry for the doubled or extended space~\cite{Blumenhagen:2011ph,Condeescu:2012sp,
Mylonas:2012pg,Deser:2016qkw,Deser:2014mxa,Alfonsi:2019ggg}. Correctly understanding in what sense the generalised vector field presented here may describe a flow, along which our canonical momentum is preserved, may well involve appealing to these constructions.



\begin{acknowledgments}

We would like to thank Dan Waldram for useful discussions.

\end{acknowledgments}





\end{document}